\documentclass[journal]{IEEEtran}
\IEEEoverridecommandlockouts

\usepackage{cite}
\usepackage{pgfplots,tikz,subcaption,varwidth,fixltx2e,paralist,graphicx,algorithmic,amsmath,amssymb,amsfonts}
\usepackage{algorithmic}
\usepackage{graphicx}
\RequirePackage{filecontents}
\usepackage{textcomp}
\usepackage{xcolor}
\usepackage{todonotes}
\def\BibTeX{{\rm B\kern-.05em{\sc i\kern-.025em b}\kern-.08em
		T\kern-.1667em\lower.7ex\hbox{E}\kern-.125emX}}
\usepackage{float}
\usepackage{url}
\usepackage{bbm}
\usepackage{lipsum,lineno}
\usetikzlibrary{backgrounds}
\usepackage{amsthm}
\usepackage[normalem]{ulem}
\usepackage{siunitx}

\newcommand{\ignore}[1]{}
\definecolor{blue-violet}{rgb}{0.54, 0.17, 0.89}
\definecolor{crimsonglory}{rgb}{0.75, 0.0, 0.2}
\definecolor{coolblack}{rgb}{0.0, 0.18, 0.39}
\definecolor{internationalkleinblue}{rgb}{0.0, 0.18, 0.65}

\usepackage{scalerel}
\usepackage{tikz}
\usetikzlibrary{svg.path}

\definecolor{orcidlogocol}{HTML}{A6CE39}
\tikzset{
	orcidlogo/.pic={
		\fill[orcidlogocol] svg{M256,128c0,70.7-57.3,128-128,128C57.3,256,0,198.7,0,128C0,57.3,57.3,0,128,0C198.7,0,256,57.3,256,128z};
		\fill[white] svg{M86.3,186.2H70.9V79.1h15.4v48.4V186.2z}
		svg{M108.9,79.1h41.6c39.6,0,57,28.3,57,53.6c0,27.5-21.5,53.6-56.8,53.6h-41.8V79.1z M124.3,172.4h24.5c34.9,0,42.9-26.5,42.9-39.7c0-21.5-13.7-39.7-43.7-39.7h-23.7V172.4z}
		svg{M88.7,56.8c0,5.5-4.5,10.1-10.1,10.1c-5.6,0-10.1-4.6-10.1-10.1c0-5.6,4.5-10.1,10.1-10.1C84.2,46.7,88.7,51.3,88.7,56.8z};
	}
}

\newcommand\orcidicon[1]{\href{https://orcid.org/#1}{\mbox{\scalerel*{
				\begin{tikzpicture}[yscale=-1,transform shape]
				\pic{orcidlogo};
				\end{tikzpicture}
			}{|}}}}

\usepackage[hidelinks]{hyperref}

\begin{document}

	\title{System-Level Dynamics of Highly Directional Distributed Networks}

	\author{\IEEEauthorblockN{
			 Asad Ali\orcidicon{0000-0001-6293-4771}, Olga Galinina\orcidicon{0000-0002-5386-1061}, and Sergey Andreev\orcidicon{0000-0001-8223-3665}}

\thanks{This work is supported by EU Marie Sk\l{}odowska-Curie (project A-WEAR $813278$) and the Academy of Finland (projects CROWN $317533$ and RADIANT $326196$).} 
	\thanks{Asad Ali, Olga Galinina, and Sergey Andreev are with Tampere University, 33100 Tampere, Finland (e-mail: asad.ali@tuni.fi;  olga.galinina@tuni.fi; sergey.andreev@tuni.fi).}
	\thanks{Manuscript received December 09, 2020; revised March 12, 2021; accepted April 05, 2021.}\vspace{-8mm}}

	\maketitle
	
	\begin{abstract}
		While highly directional communications may offer considerable improvements in the link data rate and over-the-air latency of high-end wearable devices, the system-level capacity trade-offs call for separate studies with respect to the employed multiple access procedures and the network dynamics in general. This letter proposes a framework for estimating the system-level area throughput in dynamic distributed networks of highly-directional paired devices. We provide numerical expressions for the steady-state distribution of the number of actively communicating pairs and the probability of successful session initialization as well as derive the corresponding closed-form approximation for dense deployments.
		
	\end{abstract}
	\section{Introduction}
	One of the conditions for the mass adoption of extended reality~(XR) devices is that they are made lightweight, sleek in design, and induce minimal heat dissipation, all resulting in limited processing capabilities. Hence, wireless XR units might need to offload their computationally intensive tasks to more powerful companion devices~\cite{ETSI}. To achieve a sufficient level of quality of user experience, the wireless connection between the XR unit and the computation devices should provide extreme data rate and ultra-low over-the-air latency, which cannot be supported by conventional microwave radio technologies. One of the prospective solutions to satisfy the strict XR connectivity requirements is to tap into the less congested parts of the radio spectrum, such as the extremely high frequency (EHF) and the tremendously high frequency (THF) bands. To compensate for the severe path loss degradation inherent at these frequencies, devices rely on the use of antenna arrays with a large number of elements that enable \textit{highly directional} communications and thus ensure sufficiently high antenna gains and data rates.
     		   	
	Wirelessly tethered devices of the same owner may form an independent \textit{directional on-body network}, which is likely to operate at the unlicensed spectrum in a \textit{distributed} manner. The lack of centralized coordination between highly directional wearable networks may result in increased levels of interference and degraded performance, especially in \textit{dense} and \textit{dynamic} environments. Understanding the system-level trade-offs and their asymptotic behavior is a key factor in the design of efficient and scalable XR networks, which we aim to address in this work.
	
	The system-level performance of a directional distributed network has been actively studied with methods of stochastic geometry, e.g., in terms of the link capacity~\cite{Comisso20193D} and interfering links~\cite{Kulkarni2018Correction}.  However, these works focus on performance metrics in a static system and disregard the impact of dynamic interactions between communicating entities. To capture the temporal effects of interacting nodes, queueing theory has been used widely as it provides convenient tools for modeling the system performance. In~\cite{sankararaman2017spatial}, a space-time interacting system is utilized to study spectrum sharing in ad-hoc wireless networks; the authors combine stochastic geometry and queuing theory to capture the temporal dynamics of interfering entities. 
	
	Most of the research involving spatial and temporal analysis has focused on low rate and omnidirectional transmissions. In~\cite{zhong2017heterogeneous}, stochastic geometry and queuing theory have been applied to model heterogeneous cellular networks and evaluate different scheduling policies based on successful packet delivery and dissimilar requirements of the users. At each base station (BS), packets arriving to users with heterogeneous requirements are modeled via queues. Meanwhile, in~\cite{zhong2020spatio}, authors analyze massive and sporadic access in distributed networks using a Poisson bipolar model. The packet arrival at the link is modeled as a Bernoulli process so that each transmitter is represented by an interfering queue. Although both of these models capture the spatio-temporal impact of the traffic on network performance, dynamic interactions between distributed networks of directional devices have not been addressed.
	
	In this letter, we develop a framework for assessing highly directional communications in distributed networks of advanced wearables. We utilize stochastic geometry and queueing theory tools to derive the expressions that approximate the key system-level metrics. Using asymptotic analysis, we obtain a simple closed-form expression for the average number of devices in a highly populated XR system. Finally, we evaluate the system-level performance and illustrate the effects of densification and directionality.
	\section{System Model}	
\begin{figure}[!t]
		\centering 
		\includegraphics[width=0.45\textwidth, angle=0]{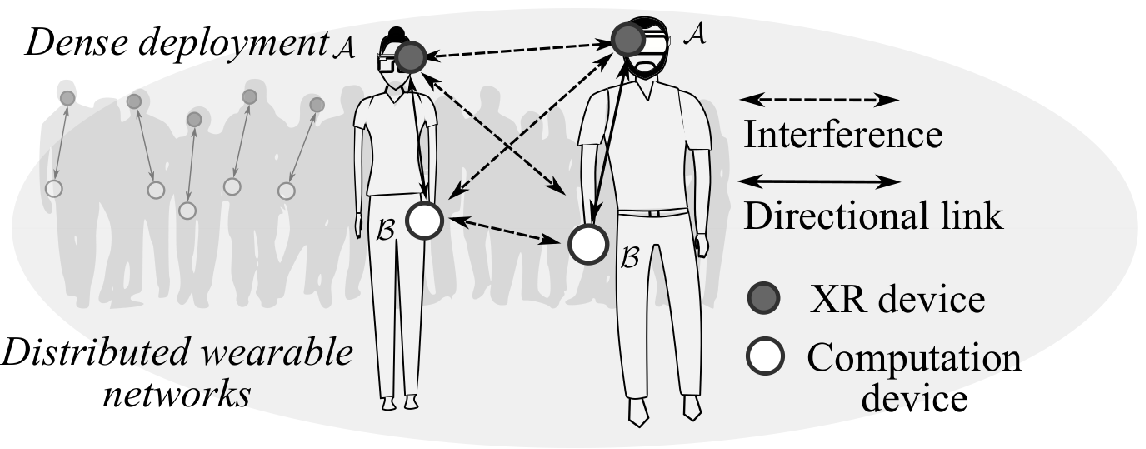}
		\caption{ Illustration for distributed networks of paired devices.}\label{fig:dep}
		\vspace{-7mm}
	\end{figure}

	\textit{Network Deployment.} We consider a \textit{distributed system}, in which paired devices $\mathcal{A}$ and $\mathcal{B}$ arrive within the area of interest $S_R$ in the 2D plane. The devices represent XR units, e.g., AR glasses or VR HMDs, and a connected computing device, such as a smartphone. The location of device $\mathcal{A}$ is distributed according to the homogeneous Poisson point process (PPP), while the paired device $\mathcal{B}$ is located around $\mathcal{A}$ within a distance of $d_{\max}$ in a random direction. We assume an arbitrary truncated distribution of distances between devices $\mathcal{A}$ and $\mathcal{B}$, $f(d),d \leq d_{\max}$.  Our scenario may represent a dense XR environment with a large number of users (i.e., concert, convention, or sports event) as depicted in Fig.~\ref{fig:dep}. 

	\textit{System Dynamics.} To capture the system dynamics, we model the arrivals of the pairs as a Poisson process with the average inter-arrival time of $1/\lambda$. The service time follows an exponential distribution with the mean of $1/\mu$. 
	During the service time, the paired devices continuously alternate between the directional transmission and the omnidirectional reception modes; thus, to the new pairs, both devices in the active pairs communicate simultaneously.
	The beams are assumed to be perfectly aligned based on, e.g., compressive sensing approach for beam management~\cite{myers2019falp}.
	
	\textit{Antenna Directivity.} The antenna directivity gain is represented as a composition of the maximum directivity $D_0$ and the directivity reduction factor $\rho(\alpha)$, which together determine the antenna radiation pattern. The directivity reduction factor is approximated by $\rho(\alpha) = 1-\frac{\alpha}{\theta}, \alpha \leq \theta$~\cite{Chukhno2019Analysis}, where $\alpha$ is the angle of deviation from the antenna boresight and $\theta$ is the half-power beamwidth (HPBW).	
	
	\textit{Radio Propagation.} The path loss is modeled according to $L(d) = C d^\kappa$, where $\kappa$ is the path loss exponent, $C$ is the propagation constant, and $d$ is the distance between the devices.  For any given transmit power $P_\mathrm{tx}$, the receive power $P_\mathrm{rx}$ is
	\begin{equation} 
	\begin{array}{c}	
		P_\mathrm{rx} = \dfrac{P_\mathrm{tx}G_{\mathrm{rx}}G_{\mathrm{tx}}}{L(d)}  =  \dfrac{P_{\mathrm{tx}}D_0}{C d^\kappa}\left(1-\frac{\alpha}{\theta}  \right),
	\end{array}\label{eq:P_rx}
	\end{equation}	 
	where $G_{\mathrm{rx}}\!=\!D_0\rho(\alpha)\!$ is the directional antenna gain and $G_{\mathrm{tx}}\!=\!1$ is the antenna gain for omni-directional antenna.   Another important parameter is the receiver sensitivity $N_{\rm thr}$ that determines the minimum received power, for which the receiver can acquire the signal. The maximum coverage distance along the antenna boresight for a given  $\!N_{\rm thr}\!$ is, therefore, expressed by {$R \!=\!\left( {\frac{P_{\mathrm{tx}} D_0 }{N_{\mathrm{thr}}C}}\! \right)^{\frac{1}{\kappa}} $}.
	
	We omit the impact of the fast fading as it has minimal effect after its compensation with standard protocol techniques, e.g., forward error correction (FEC). The slow fading can be modeled by introducing an additional random variable to the considered propagation model, which leads to $\!P_\mathrm{rx}\!=\!\frac{P_{\mathrm{tx}}D_0 \Omega}{C d^\kappa}\left(1\!-\!\frac{\alpha}{\theta}\right)\!$, where $\Omega$ follows, e.g., a log-normal distribution. The randomness of $P_\mathrm{rx}$ would affect the coverage distance $R$ and the interference between the pairs, which brings an additional degree of complexity to further analysis. According to our numerical experiments, however, fading has a marginal effect on the average system-level performance and, hence, we omit the consideration of small-scale fading for the sake of better tractability. 

	\textit{User Admission.} In our model, a new pair is accepted if both devices do not face excessive interference from any of the actively communicating pairs in the system. Given $n$ active pairs already in service, the admission of the $(n\!+\!1)$-th pair is allowed only if the following condition is satisfied 
	
	\begin{equation}
	\begin{array}{c}
		P_\mathrm{rx}^{\mathcal{A}_{i}/\mathcal{B}_i\to \mathcal{A}_{n+1}/\mathcal{B}_{n+1}}< N_{\rm thr} \ \ \forall \ \  i\!=\!1,...,n, 
	\end{array}
	\end{equation} 
	where $P_\mathrm{rx}^{\mathcal{A}_{i}/\mathcal{B}_i\to \mathcal{A}_{n+1}/\mathcal{B}_{n+1}}$ implies the received power at either $\mathcal{A}_{n+1}$ or $\mathcal{B}_{n+1}$   from $\mathcal{A}_{i}$ or  $\mathcal{B}_{i}$. Hence, for a successful admission, the received power at either device of the $(n+1)\text{-th}$ pair from all devices $\mathcal{A}_i$ and $\mathcal{B}_i$ of all active $n$ pairs should be less than the receiver sensitivity. These admission criteria are based on the Distributed Coordination Function (DCF) of IEEE 802.11, which allows transmission \textit{if and only if} a channel is observed to be idle. We note that the considered criteria lead to a hard-core point process for the spatial distribution of active users. 
	\begin{figure}[!t]
		\centering 
		\includegraphics[width=0.28\textwidth, angle=0]{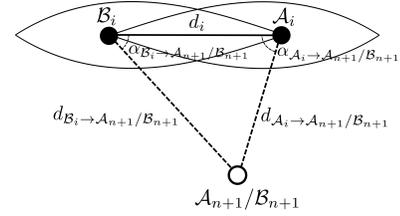}
		\caption{ Illustration for one pair of communicating devices.}\label{fig:dis}
		 \vspace{-6mm}
	\end{figure}

	To analyze the system behavior, we propose a mathematical framework in the following section.
	\section{Proposed Analysis} 
	\subsection{State Aggregation}	
\begin{figure}[!t ]
		\vspace{-3mm}
		\centering 
		\includegraphics[width=0.38\textwidth, angle=0]{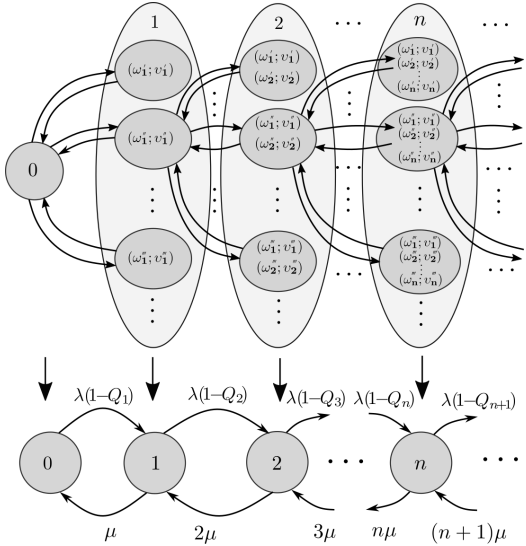}
		\vspace{-1mm}
		\caption{ Illustration of state aggregation approach.} \label{fig:queue_big}	\vspace{-6mm}
	\end{figure}

	Our dynamic system can be modeled as a Markov process $S(t)$, where the state is determined by the number of active pairs within the area of interest and their locations as $[n|(w_1;\!v_1),\!(w_2;v_2),\!\cdots\!,(w_n;v_n)]$, where $w_i$ and $v_i$ correspond to the locations of the devices in the $i$-th pairs, as shown in Fig.~\ref{fig:queue_big} (``$\prime$" indicates different locations). For $i$ active pairs in the system, the number of possible locations that devices can occupy is infinite, and, thus, the number of states is uncountable, which results in a high complexity of the mathematical model.
		
	For simplification, we employ the state aggregation approach~\cite{andreev2014analyzing}, where the states with $n$ active pairs are unified into one state $n$, regardless of the location of each device, as shown in Fig.~\ref{fig:queue_big}. While the state transitions of the resulting process $S^\prime(t)$ depend only on the current state, we incorporate the information about the previous admissions into the state transition probabilities by introducing probabilities $Q_n$, which are detailed further. The resultant Markov process $S^\prime(t)$ is a birth-death process as the state $n$ changes by~$\pm 1$.	
	\subsection{Queueing Model}  
	The state distribution of process $S^\prime(t)$ is given by
	\begin{equation}\label{eq:pi_n}
	\pi_m = \pi_0 \left(\dfrac{\lambda}{\mu} \right)^m \dfrac{\prod_{n=1}^{m-1} (1-Q_n)}{m!}	, m \geq 1,
	\end{equation} 
	where $Q_n = \Pr\{(n+1)\textit{-th rejected} |n \textit{ active pairs}\}$ is the probability for a new pair to be rejected when there are $n$ active pairs already running in the system; $\pi_0 \!=\! \left(1+\sum_{m=1}^{\infty} \left({\lambda/}{\mu} \right)^m {\prod_{n=1}^{m-1} (1\!-\!Q_n)}/{m!}\!  \right)^{-1}$ follows from the normalization condition. Based on the steady-state distribution, one can derive the key metrics characterizing our system, i.e., the average number of pairs and the acceptance probability:  	
	\begin{equation}
	\mathrm{E}[\mathrm{N}] = \sum_{n=1}^{\infty} n \pi_n \text{ and }
	\mathrm{P}_\mathrm{accept} = \pi_0 + \sum_{n=1}^{\infty}(1-Q_{n})\pi_n. 
	\end{equation}	
	
	In what follows, we calculate the probability $Q_{n}$. A new pair is rejected by the system when it encounters strong interference from any of the active pairs. For state $n=0$ (empty system), the newly arriving pair encounters no interference; thus, it is always accepted into the system. For states $n>0$, acceptance or rejection of a pair depends on the number of already running pairs. 
	When the $(n+1)$-th pair emerges in the coverage area of any of the active pairs, the received power at the $(n+1)$-th pair is greater than $N_{\mathrm{thr}}$, and the new pair cannot be admitted. The probability of this event is given by $\frac{S_{\mathrm{cov}}}{S_R}$, where $S_R$ is the area of interest and $S_{\mathrm{cov}} \simeq  \frac{2 R^2\kappa \theta}{(2+\kappa)}$ is the area covered by the antenna radiation of both devices in the pair (the expression for $S_{\mathrm{cov}}$ is detailed in Appendix). Based on these considerations, we formulate the following proposition. \\
	\textbf{Proposition 1.} The probability of $(n+1)$-th pair being rejected upon arrival can be approximated as 
	\begin{equation}\label{eq:Qn}
	Q_{n} = \min\left(n \left(\dfrac{ 2R^2\kappa \theta}{S_{\mathrm{R}}(2+\kappa)}\right),1\right) = \min( n \gamma ,1),
	\end{equation} 
	where $\gamma =  \frac{ 2R^2\kappa \theta}{S_{\mathrm{R}}(2+\kappa)}$. Here, $Q_{n}$ is modeled as a piece-wise function; however, one may observe that in reality, $Q_{n}$ is nonlinear. We further approximate $Q_{n}$ with a generalized logistic function, which is widely used for modeling population growth.
	
	\textbf{Proposition 2.} The approximation of $Q_{n+1}$ by the generalized logistic function is given as 
	\begin{align}\label{eq:1Q_prime}
	\tilde Q_{n} & = \dfrac{2}{1+e^{-2n\gamma }}-1 .			
	\end{align}		
	\begin{proof}
		 We consider logistic function $z(n)= \frac{2}{1+e^{-b\cdot n }}-1$, where 
		the coefficients follow from conditions $ {z(n) \xrightarrow[\!\!n\! \rightarrow \!\infty\!\!]{} 1} $\ignore{ $ \underset{n \rightarrow \infty}{z(n) \rightarrow 1} $ } and $ z(0) =0 $. We estimate the remaining parameter $b$ as follows. Assuming that at smaller $n$ the slope of $z(n)$ repeats the slope of $Q_{n}$ given by $\gamma$, we find the derivative ${\left. \frac{\partial }{\partial n} \left(z(n)\right) \right|_{n=0}= \frac{2be^{-b\cdot n }}{(1+e^{-b\cdot n })^2}	= \frac{b}{2}}$. We approximate the parameter $b$ by $2\gamma$, and the resultant expression immediately yields \eqref{eq:1Q_prime}.
	\end{proof}
	 In our model, the acceptance probability is inversely proportional to the number of pairs in the system, i.e., as the deployment densifies, the probability of acceptance reduces, thus exhibiting a self-limiting behavior of the population growth.

	\textbf{Theorem 1.} We may further approximate $Q_{n}$ as
	\begin{equation}\label{eq:Q_prime}
	\tilde Q_{n} = 1- e^{-2n\gamma}.
	\end{equation}
	\begin{proof}
		Derivation of (7) is detailed in Appendix.
	\end{proof}
	Numerical calculation of the steady-state distribution using~(\ref{eq:pi_n}) may appear computationally intensive for large numbers. For the purposes of asymptotic analysis, we derive the following expression, which is further used to approximate the average number of pairs:
	\begin{equation}
		\begin{array}{c}
		\tilde \pi_m = \tilde \pi_0 \left(\dfrac{\lambda}{\mu}\right)^m \dfrac{e^{-\gamma m(m-1)} }{ m!},
		\end{array}\label{eq:pi_n2}
	\end{equation}	  
	where $\tilde	\pi_0 =\left(1+\sum \limits_{m=1}^{\infty}  \left(\dfrac{\lambda}{\mu}\right)^m \dfrac{e^{-\gamma m(m-1)} }{ m!} \right)^{-1}$. \vspace{1mm}
	
	\textbf{Theorem 2.}
	Based on the above approximation for the steady-state distribution, we obtain a closed-form expression for the average number of active pairs in the system as
	\begin{align} \label{eq:En3}
		E[N] = \frac{1}{2 \gamma} W\left( 2  \gamma \dfrac{\lambda}{ \mu }e^{\gamma}\right) ,
	\end{align}
	 where $W(y)$ is the Lambert W function, a solution to~$xe^x=y$.
	\begin{proof}
		Derivation of (\ref{eq:En3}) is detailed in Appendix. 
	\end{proof}

	\section{Numerical Results}
	We consider a deployment where high-end wearable devices are wirelessly paired with the corresponding computation units within a circular area of interest having radius $R_{d}=3000$, $S_R=\pi R_{d}^2$. The service rate is set at $\mu=1$\,s$^{-1}$ for an active session of the pair. The envisioned system operation follows the principles of communication protocols for distributed systems such as IEEE 802.11ad/ay, with a particular focus on the carrier sensing procedures for multiple access. Furthermore, we assume the line of sight operation, according to the Friis transmission equation with $\kappa=2$ and \ignore{$C=(4 \pi f/c)^2$} $C=6.3 \times 10^6$, where $f=60$\,GHz. The maximum antenna directivity is approximated as  $D_0 = {2}/\left({1 - \cos \frac{\theta}{2}}\right)$. For all devices, the receiver sensitivity is set at $N_\mathrm{thr}=-78$\,dBm.
	\subsection{Validation of Proposed Approximation}
	First, we validate our approximation provided in (\ref{eq:pi_n2}) and (\ref{eq:En3}) with Monte Carlo simulations. For more realistic modeling of our wearable deployment, the paired devices are randomly placed in a 3D volume (represented by a cuboid of size 0.3\,m$\times$0.5\,m$\times$0.6\,m). We consider projections of the paired devices in the 3D space onto the horizontal 2D plane. By doing so, we mimic 2D beamforming between the paired devices. For the analytical modeling, the 2D distance $d$ between paired devices is replaced with the expected value of distances, $E[d]$, where the paired devices are distributed in the cuboid. In the simulations, a new pair checks the interference two ways, i.e., it neither experiences nor causes excessive interference.
	
\begin{figure}[t!]
		\vspace{-1mm}
		\centering 
		\includegraphics[width=0.441\textwidth, angle=0]{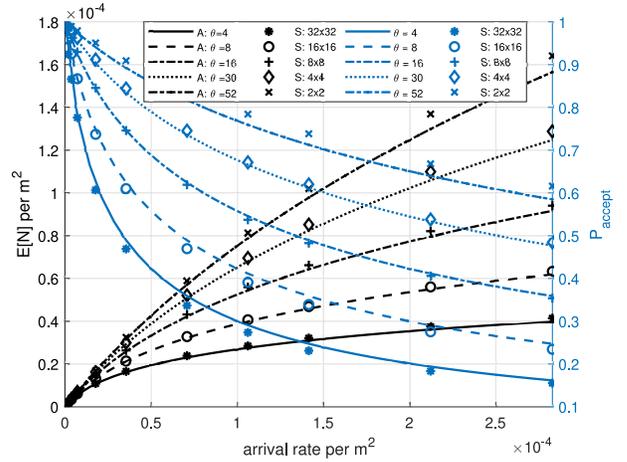}
		\vspace{-1mm}
		\caption{$E[N]$ per m$^2$ and acceptance probability, $\mathrm{P}_{\mathrm{accept}}$, vs. arrival rate, $\lambda$, per m$^2$, for varied HPBW and antenna arrays; $P_\mathrm{tx}\!=\!10$\,dBm and $N_\mathrm{thr}\!=\!-78$\,dBm. Here, “A” stands for analysis and “S” stands for simulation.}\label{fig:lambda_EN}\label{fig:beam1}	\vspace{-3.5mm}
	\end{figure}

	In Fig. \ref{fig:beam1}, we illustrate the impact of the arrival rate per m$^2$ on the average number of active pairs, $E[N]$, per m$^2$ and the pair acceptance probability, $\mathrm{P}_\mathrm{accept}$. For the Monte Carlo simulations, we utilize the radiation patterns of uniform rectangular phased arrays, whose HPBW corresponds to the HPBW used in the analysis (i.e., $2 \times 2$ corresponds to $\theta=\ang{52}$). The antenna array radiation patterns are produced by the MATLAB Sensor Array Analyzer toolbox and projected onto the 2D plane. The simulation and analytical results are largely in agreement with one another for both metrics of interest. 
	
	We note that our analytical expressions provide results consistent with the simulation data for a wide range of parameters (i.e., $\!\!P_\mathrm{tx}\!\!=\!\! -20\!\text{ to }\!20\!$ dBm and $\theta\!=\ang{4} \!\text{ to  }\!\ang{52} $) if the area of interest is large enough to compensate for the excessive radiation that falls outside of it. A slight saturation in $E[N]$ can be observed at the increased arrival rate, which becomes more evident for the curves with narrower HPBW $\theta$. On the contrary, the acceptance probability, $\mathrm{P}_\mathrm{accept}$, declines for the increased arrival rate due to the growth in the number of active pairs, which act as potential interferers. This behavior can be explained by the self-limiting nature of such systems. 
	
	The results presented in Fig. \ref{fig:beam1} stem from the 2D radiation pattern due to the 2D beamforming consideration. In the case of 3D beamforming, the system might demonstrate different behavior. In particular, with the 3D beamforming, the system can benefit from steering antennas in both horizontal and vertical planes, which allows the interference from the neighbors to also be dispersed in the vertical plane, whereas with the 2D beamforming, the interference is concentrated mostly in one plane. We note that our results apply to the case of a single channel; in the case of multiple channels, there should be separate $E[N]$ and $\mathrm{P}_\mathrm{accept}$ for each one.	
	\subsection{Optimization of System Throughput}
\begin{figure}[t!]
		\vspace{-3mm}
		\centering 
		\includegraphics[width=0.441\textwidth, angle=0]{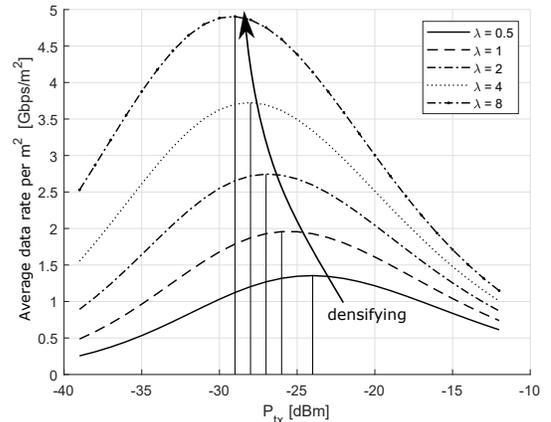}
		\vspace{-2mm}
		\caption{Average data rate per m$^2$ vs. transmit power, $\!P_\mathrm{tx}$, for varied arrival rate, $\!\lambda$,  per m$^2$ and fixed HPBW $\!\theta\!\!=\!\!\ang{30}$; $N_{\mathrm{thr}}\!=\!-\!78$\,dBm corresponds to MCS0, and $\mathrm{SNR}_{\max}\!\!=\!\!20$\,dB. Data rate can be maximized by adjusting $P_\mathrm{tx}$ according to density. }\label{fig:beam2}\vspace{-3mm}
	\end{figure}

\begin{figure}[h!]
		\vspace{-3mm}
		\centering 
		\includegraphics[width=0.441\textwidth, angle=0]{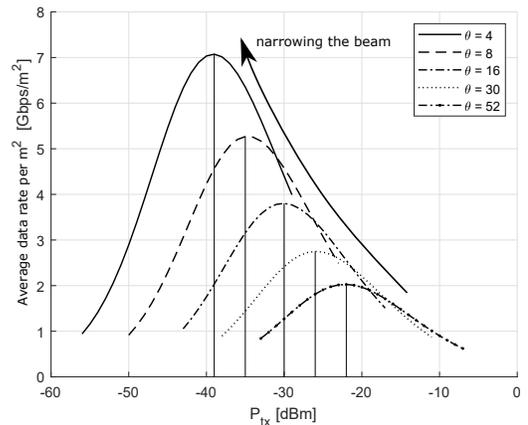}
		\vspace{-2mm}
		\caption{Average data rate per m$^2\!$ vs. transmit power, $\!P_\mathrm{tx}$, for varied HPBW, $\theta$;  arrival rate per m$^2\!\!=\!$  $\!2$~\cite{weppner2013bluetooth}, $\!N_{\mathrm{thr}}\!\!=\!-\!78$\,dBm corresponds to MCS0, $\mathrm{SNR}_{\max}\!\! =\! 20\! $ dB. Data rate can be maximized for each HPBW by choosing appropriate $P_\mathrm{tx}$.  }\label{fig:beam3}\vspace{-6 mm}
	\end{figure}

	In Fig. \ref{fig:beam2} and \ref{fig:beam3}, we illustrate the impact of the transmit power on the achievable data rate in a dense environment. The average data rate is estimated by Shannon-Hartley theorem as
	\begin{equation}
		c= B \log_2 \left(1+\min\left(\frac{P_\mathrm{rx}}{P_\mathrm{n}},\mathrm{SNR}_{\max}\right)\right),
	\end{equation}
	where $P_\mathrm{n}$ is the noise power and $\mathrm{SNR}_{\max}$ is the SNR determined by the maximum modulation and coding scheme (MCS). Due to the dense deployment, one may utilize the rule of ``K closest neighbors'' to capture interference from the neighbors by assuming $P_\mathrm{n}=N_{\mathrm{thr}}\times$K with $\text{K}=6$.
	
	
	In Fig. \ref{fig:beam2}, we plot the average data rate per m$^2$ against the transmit power, $P_\mathrm{tx}$, for different arrival rates. We observe that there exists an optimal transmit power, for which the maximum data rate can be attained. As the system densifies, higher rates per m$^2$ can be reached with lower power. In Fig. \ref{fig:beam3}, we study the dependence of the average data rate per m$^2$ on the transmit power, $P_\mathrm{tx}$, for varied $\theta$. For narrower beams, even though the number of active pairs is reduced, the system can offer higher area throughput due to better link performance. However, that occurs only for certain combinations of HPBW and $P_\mathrm{tx}$ values, i.e., for any fixed HPBW, there exists an optimum value of the transmit power that maximizes the achievable data rate. We note that lower values of the optimal power might lead to a decrease in the link throughput.  To control the range of the individual data rate, one should add corresponding constraints to the considered optimization problem, where objective function $c E[N]$ is a function of $P_\mathrm{tx}$. 
	
	 The interplay between the transmit power and the system-level interference depends on the beamwidth and the arrival rates. While in a practical distributed system, optimizing $P_\mathrm{tx}$ is relatively complicated, our analysis can advise on the choice of optimal $P_\mathrm{tx}$ if the medium access protocols incorporate Transmit Power Control (TPC) mechanisms, such as those in IEEE 802.11ay~\cite{chen2019Millimeter}.
	
	\section{Conclusion}
	In this letter, we proposed a mathematical framework to conduct the system-level analysis for distributed networks of wearables based on highly directional communications. We presented a novel numerical solution for the steady-state distribution and the acceptance probability as well as derived a closed-form expression for the average number of pairs using asymptotic analysis. We validated our proposed model with Monte Carlo simulations for realistic antenna patterns while considering projections of paired devices onto the 2D plane, thus, mimicking the 2D beamforming. Based on these numerical results, we observed that the data rates can be maximized by adjusting the transmit power according to the beamwidth and the density of devices. Our presented analytical expressions can become a useful reference for further performance optimization of highly-directional dense wearable systems.

	\appendix
	\subsection{Calculation of the Area of the Beam}
\begin{figure}[h!]\vspace{-2mm}
		\centering 
		\includegraphics[width=0.30\textwidth, angle=0]{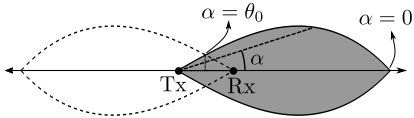}
		\vspace{-1mm}
		\caption{Illustration of beam coverage area.}\label{fig:beam_coverage}
		\vspace{-1mm}
	\end{figure}

	Here, we derive an analytical expression for the beam coverage area (Fig.~\ref{fig:beam_coverage}). For deviation angle $\alpha$, distance $d(\alpha)$ between the antenna and the border of the coverage area is 	
	\begin{equation}
		\begin{array}{c}
	d(\alpha)  = \left( {\frac{ P_{\mathrm{Tx}} D_0 \rho( \alpha ) }{P_{\mathrm{thr}} C }} \right)^{\frac{1}{\kappa}} \!= 
		R \left(  {1-\dfrac{\alpha}{\theta}} \right)^{\frac{1}{\kappa}} \!\!\!\!, \text{ if } \alpha \leq \theta.   
		\end{array}
	\end{equation}
	The \textit{x}- and \textit{y}-coordinates of arbitrary points at the border are 
	$x(\alpha)   = d(\alpha) \cos(\alpha)  =   R \left(  {1-{\alpha}/{\theta}} \right)^{\frac{1}{\kappa}}\cos (\alpha)  , \text{ if }\alpha \leq \theta$ and 
	$y(\alpha)   = d(\alpha) \sin(\alpha)  =   R \left(  {1-{\alpha}/{\theta}} \right)^{\frac{1}{\kappa}}\sin (\alpha)  , \text{ if }\alpha \leq \theta$.

	To find the area of the beam, we use the integral expression
	\begin{align}
		A_{\text{beam}}  = 2 \int_{0}^{\theta_0}  x(\alpha)  y'(\alpha) \mathrm{d}\alpha   =	\dfrac{ R^2\kappa \theta_0}{(2+\kappa)}.
	\end{align}
	In our analysis, we disregard the overlap between two beams and replace $\theta_0$ by $\theta$ for simplicity. Therefore, the full coverage area of a pair can be expressed as $S_{\mathrm{cov}}\simeq 2  A_{\text{beam}} \simeq \frac{ 2 R^2\kappa \theta}{(2+\kappa)}$.

	\subsection{Proof of Theorem 1}
	According to Proposition 1, 
	$\tilde Q_{n} ={2}/\left( {1+e^{-2n\gamma }} \right)-1   . $ 
	Hence, for large $n$, the pair acceptance probability $(1 - \tilde Q_{n})$ can be approximated as		
	\begin{equation}
		\begin{array}{c} \label{eq:prod}
		(1 - \tilde Q_{n}) =   \dfrac{2e^{-2n\gamma }}{1+e^{-2n\gamma }} \approx e^{-2n\gamma },       			
		\end{array}
	\end{equation}
	which can also be illustrated via a numerical comparison. Furthermore, by performing $\prod_{n=1}^{m-1}$ operation over $(1-Q_{n})$, as required in (\ref{eq:pi_n}), we obtain 
	$\prod_{n=1}^{m-1}	(1-Q_{n})	 =    e^{-\gamma m(m-1) }$ .
	\subsection{Proof of Theorem 2}
	At higher loads, the discrete steady-state distribution can be fitted with a bell-shaped curve, and, thus, its maximum approximately corresponds to $E[N]$. Hence, assuming continuous $n$, we estimate the maximum using $\frac{\partial \tilde \pi_n}{\partial n}\!$ as 
	\begin{equation}
		\begin{array}{cc}
		\! \! \! 	\frac{\partial \tilde \pi_n}{\partial n}\! =&   \tilde\pi_0\! \frac{\partial}{\partial n}\! \! \left(\! \!  \left(\! \dfrac{\lambda}{\mu}\right)^n \! \! \! \! \dfrac{e^{-\gamma n(n-1)} }{ \sqrt{2 \pi n} \left(\! \dfrac{n}{e}\! \right)^n}\! \!  \right) \! \! = \! \dfrac{\tilde \pi_0 \left(\dfrac{\lambda}{\mu}\right)^n  e^{-\gamma n(n-1)+n}}{2^\frac{3}{2} n^\frac{2n+3}{2} \sqrt{\pi}} \nonumber \\
		& \ \ \ \ \ \ \ \times \biggr(\!\!  2n\biggr(\! \! \log\left(\! \dfrac{\lambda}{\mu}\! \right)\! -\!\gamma (2n\!-\!1\!)  \!-\!\log (n)\! \!\biggr)\!-\! 1\!\biggr)\! .\! \! \! \! %
		\end{array}
	\end{equation}
	The solution to equation $\frac{\partial \tilde \pi_n}{\partial n}\! =0$ is $n^* = \frac{1}{2 \gamma} W\left( 2  \gamma \dfrac{\lambda}{ \mu }e^{\gamma}\right)$, where $W(\cdot)$ is the Lambert W function. Therefore, we can approximate $E[N]$ as	
	$E[N] = \frac{1}{2 \gamma} W\left( 2  \gamma \dfrac{\lambda}{ \mu }e^{\gamma}\right)$.


	\bibliographystyle{IEEEtran}
	\bibliography{refs_letter}
	
\end{document}